Comment on "Symmetry energy and the isospin dependent equation of state"

A. Ono[1,2], P. Danielewicz[1], W.A. Friedman[3], W.G. Lynch[1], M.B. Tsang[1]

[1]National Superconducting Cyclotron Laboratory and Department of Physics and Astronomy, Michigan State University, East Lansing, MI 48824, USA
[2]Department of Physics, Tohoku University, Sendai, Japan,
[3]Department of Physics, University of Wisconsin, Madison, WI 53706, USA.

## Abstract

In a recent paper, Shetty et al [Phys. Rev. C 70, 011601R (2004)] reported that the experimental isoscaling parameters $\alpha_{exp}$ from several reactions favor the Gogny-AS effective nucleon-nucleon interaction over Gogny interaction. This conclusion is reached by comparing data to Antisymmetrized Molecular Dynamics (AMD) predictions for collisions of calcium isotopes. The specific simulations produce excited fragments at t=300 fm/c before they decay. Sequential decay calculations of the excited fragments suggest that the isoscaling parameter $\alpha_{pri}$ from the AMD calculations could be reduced by as much as 50%. We also explore the linear equation used to relate other reactions to the AMD calculations for Ca+Ca collisions. The uncertainty in the relation is larger than the differences between the predicted $\alpha_{pri}$ when using Gogny and Gogny-AS interactions.

The density dependence of the asymmetry term is a fundamental property of bulk nuclear matter. It plays an important role in the structure and stability of neutron stars and in the dynamics of their formation in type II supernovae. In a recent paper [1], Shetty et al. compared experimental data for fragment isotopic distributions to Anti-symmetrized Molecular Dynamics calculations (AMD) [2] and drew conclusions about the density dependence of the asymmetry term of the nuclear equation of state. These conclusions rely on specific comparisons between the AMD predictions for the calculated isotopic yields of primary fragments obtained in head-on collisions of $^{40}Ca+^{40}Ca$, $^{48}Ca+^{48}Ca$, and $^{60}Ca+^{60}Ca$ [2] at t=300 fm/c to yields of ground state fragments measured in a number of reactions. The experimental data include collisions of $^{58}Fe+^{58}Fe$, $^{58}Fe+^{58}Ni$, $^{58}Ni+^{58}Ni$, $^{124}Sn+^{64}Ni$, $^{112}Sn+^{58}Ni$ as well as p and He induced reactions on $^{112}Sn$ and $^{116}Sn$ [1]. (In another similar study, the same authors have also included the collisions of $^{40}Ar+^{58}Fe$, $^{40}Ar+^{58}Ni$, $^{40}Ca+^{58}Ni$ [3]). To make this comparison, Shetty et al. assume (i) that the yield ratios of ground state fragments can be directly compared to the calculated yield ratios of primary fragments and (ii) that the asymmetry of the primary fragments from different reactions can be extrapolated from the AMD predictions of calcium isotope reactions at b=0 fm. We test the validity of these assumptions and find that (i) the sequential decay effects from hot fragments are substantial and (ii) the method of extrapolating the asymmetry of the primary fragments proposed in [1] is problematic. This implies that the conclusions reached in ref. [1] regarding the density dependence of the asymmetry term are premature.

To put these statements into context, we introduce the isoscaling parameters used both in the interpretation of the AMD calculations and in the analysis reported in ref. [1]. Specifically, calculations [2,4] and experiments [5] have demonstrated that the isotopic yield ratios, for two similar reactions that differ only in the proton fraction, depend exponentially on the proton (Z) and neutron (N) number of an emitted fragment,

$$Y_2(N,Z)/Y_1(N,Z) \propto e^{\alpha N+\beta Z} \qquad (1)$$

where $Y_i(N,Z)$ is the isotope yield from reaction i. By convention, reaction 1 refers to the less neutron-rich system of the two.

In the context of AMD simulations for central collisions of Ca isotopes at E/A=35 MeV [2,6], we demonstrated that the calculated yields of excited fragments satisfy the relationship

$$\alpha_{pri} = \frac{4C_{sym}}{T}\left[\left(\frac{Z_1}{A_1}\right)^2 - \left(\frac{Z_2}{A_2}\right)^2\right] = \frac{4C_{sym}}{T}[\Delta(\frac{Z}{A})^2], \qquad (2)$$

where $\alpha_{pri}$ is the isoscaling parameter extracted from the calculated yields of primary fragments before secondary decay, $(Z/A)=(Z/A)_{liq}$ is the average proton fraction of the fragments (referred to as liquid) and $C_{sym}$ is the symmetry energy of uniform nuclear matter at a reduced density. The predicted values for $\alpha_{pri}$ depend on the density dependence of the asymmetry term used in the simulations. The excited fragments produced in the AMD simulations at t=300 fm/c typically have excitation energies of E*/A≈3 MeV. Thus, calculations of the secondary decay are needed to predict the ground state yields that are experimentally measured. Since the focus of calculations in ref. [2] was on the isospin dynamics and not on a comparison to data, sequential decay calculations were not carried out.

In Ref. [1], theoretical yield ratios [2] for the excited fragments are compared directly to experimentally measured ground state yield ratios. To be valid, such an approach requires that $\alpha_{pri}$ values obtained before secondary decay be approximately equal to the corresponding values for $\alpha_{final}$ after secondary decay. To assess the accuracy of this approach, we have calculated $\alpha_{final}$ using a secondary decay code, MSU Decay [7] that incorporates a very large table of the available empirical information such as the binding energies, spins, isospins, parities and decay branching ratios for isotopes with Z ≤ 15. For heavier fragments with Z > 15 and for fragments not included in the table, the Monte Carlo decay code, Gemini [7,8] is employed.

The range of asymmetries of the experimental systems in ref [1] is mainly confined to the asymmetries between $^{48}$Ca+$^{48}$Ca (reaction 2) and $^{40}$Ca+$^{40}$Ca (reaction 1) systems. The dashed lines in Figure 1 represent the linear interpolation for $\alpha_{pri}$ before secondary decay [2], as a function of $\Delta(Z/A)^2_{liq}$. The left panel shows the predictions for the Gogny interaction and the right panel shows the predictions for the Gogny-AS interaction. The solid lines in the figure represent the linear interpolation for $\alpha_{final}$ from these two systems after secondary decay. These comparisons suggest that isoscaling parameter values may decrease by 50-60% when sequential decay is taken into account. Essentially the same results are obtained using a different sequential decay code based on [9]. Sequential decay effects of similar magnitude have been reported by Kowalski et al., for AMD simulations of Zn induced reactions on Ni, Mo and Au [10, 11]. Somewhat larger reductions of $\alpha_{pri}$ due to secondary decays have also been found in the dynamical Stochastic Mean Field simulations of Sn+Sn collisions [12].

We note that the secondary decay corrections to dynamical simulations are larger than the corresponding corrections to equilibrium statistical model predictions [13]. The difference may arise from the different conditions encountered in current statistical and dynamical models, which lead to the production of different ensembles of excited fragments. One therefore cannot use the ensembles of excited fragments from a statistical multifragmentation model in place of the AMD ensembles to prove that the sequential decay corrections to isoscaling parameters calculated from the AMD ensembles are negligible. If one calculates the sequential decay of the AMD ensembles, all present calculations of the secondary decay corrections to our AMD simulations exceed the difference between the $\alpha_{pri}$ obtained in AMD calculations for Gogny and Gogny-AS interactions [1].

To compare data to calculations, Shetty et al. proposed a linear relationship between the proton fraction of the liquid and the proton fraction of the initial system, $\Delta(Z/A)^2_{liq} = f \cdot \Delta(Z/A)_o^2$, where f=0.486 and 0.527 for Gogny and Gogny-AS effective interactions, respectively for many different types of reactions (Figure 3 in ref. [1]). These f values are obtained from $^{40}Ca+^{40}Ca$, $^{48}Ca+^{48}Ca$ and $^{60}Ca+^{60}Ca$ collisions at b=0 fm at E/A=35 MeV and may not be applicable to other reactions under consideration. In fact, for central and semi-central collisions (b<8 fm), AMD simulations with Gogny interactions of $^{64}Zn+^{197}Au$, $^{64}Zn+^{92}Mo$, $^{64}Zn+^{58}Ni$ systems at E/A=35 MeV yield a value of f~0.75 [10,11]. Without carrying out exact simulations for every reaction, we cannot estimate further the variations of f over the reactions studied by Shetty et al. We use the horizontal arrows in figure 1 to indicate that the x-values of the data points are not well determined without more simulations. The uncertainties in the constant f (>50%), however, are comparable to or larger than the differences between the Gogny and Gogny-

AS predictions (<25%) that Shetty et al. want to exploit in order to determine the asymmetry term of the EOS.

If the density and the temperature of the fragment formation depend on the conditions and types of reactions as suggested in the simulations of the $^{64}$Zn induced reactions, $C_{sym}/T$ may vary with systems. Then, it may be inappropriate to compare the experimental data points from many different reactions to the lines in Figure 1 obtained from the central collisions of the Ca isotopes. Even though it is desirable to find a general relationship to link results from different experiments, one can only achieve such goals after careful studies. For example, one should verify the reliability of the proposed method by doing simulations with a range of reactions at non-zero impact parameters and understand how f varies with impact parameter before asserting that dependence of the calculations on impact parameter or other quantities can be neglected.

In summary, we find that the two approximations used in ref. [1], ie. the neglect of secondary decay and the simple scaling of reactions according to the initial proton fraction differences, introduce excessive errors in the comparison between theory and experiment. Consequently, the procedure used in ref. [1] to extract symmetry energy and the conclusions that fragment data show preference for the Gogny-AS interaction is not justified.

This work was supported by the National Science Foundation under Grant Nos. PHY-01-10253, PHY-0245009, by the Japan Society for the Promotion of Science and the US National Science Foundation under the U.S.-Japan Cooperative Science Program (INT-01-24186).

Fig. 1: (Color online) Isoscaling parameter $\alpha$ as a function of $\Delta(Z/A)_{liq}^2$. The dashed (solid) lines represent linear interpolated results from primary (final) fragments yields calculated in $^{40}Ca+^{40}Ca$ and $^{48}Ca+^{48}Ca$ collisions at E/A=35 MeV and b=0 fm [2]. The left and right panels represent calculations using the Gogny and Gogny-AS interactions. Symbols are experimental data points taken from different reactions [1]. The horizontal arrows indicate that the abscissa values for the data points are poorly determined (see text for details).

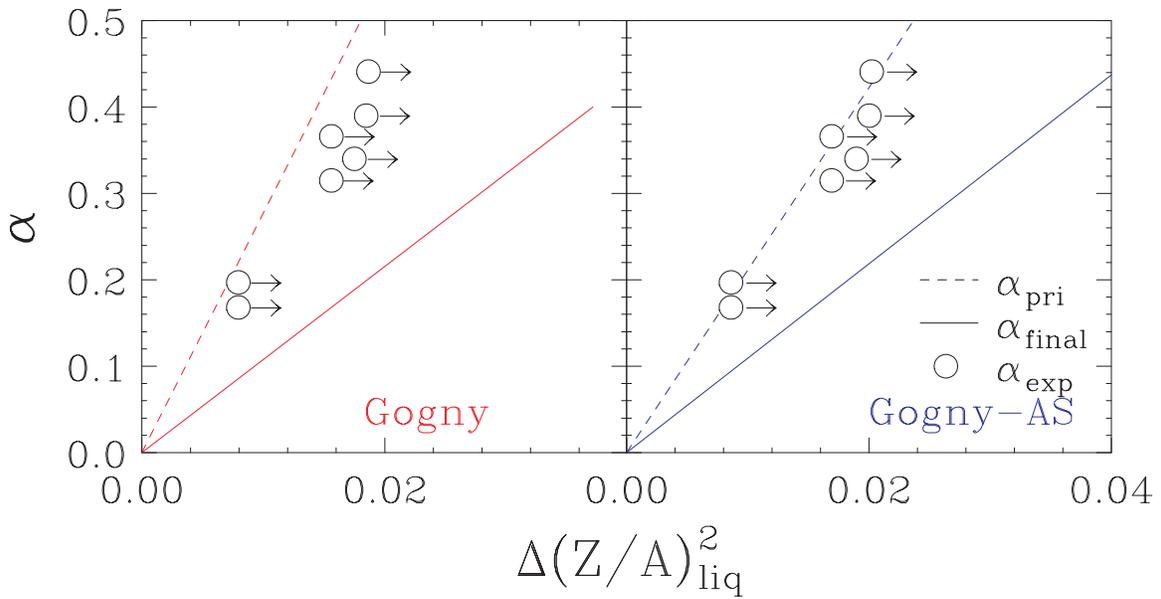